\documentclass[cameraready]{Interspeech}
\title{BabyHuBERT: Multilingual Self-Supervised Learning\texorpdfstring{\\}{ }for Segmenting Speakers in Child-Centered Long-Form Recordings}

\author[affiliation={1}, orcid=0009-0004-6414-3964, equalcontribution, correspondingauthor]{Théo}{Charlot}
\author[affiliation={1}, orcid=0009-0009-2158-597X, equalcontribution, correspondingauthor]{Tarek}{Kunze}
\author[affiliation={1}, orcid=0000-0002-9377-9150]{Maxime}{Poli}
\author[affiliation={1}, orcid=0000-0003-2979-4556]{Alejandrina}{Cristia}
\author[affiliation={1}, orcid=0000-0002-7814-2952]{Emmanuel}{Dupoux}
\author[affiliation={2}, orcid=0000-0002-6005-9368]{Marvin}{Lavechin}

\address{
    $^1$ LSCP, DEC, ENS, EHESS, CNRS, PSL University, France \\
    $^2$ Laboratoire d'Informatique et Systèmes, Université Aix-Marseille, CNRS, France
}

\email{theo.charlot@ens.psl.eu}


\usepackage{comment}

\newcommand{\class}[1]{\small \texttt{#1}}

\newcommand{\LENA}{LENA\texttrademark~}

\usepackage{xcolor}
\usepackage{hyperref}
\hypersetup{
    colorlinks,
    linkcolor={red!50!black},
    citecolor={blue!60!black},
    urlcolor={red!50!black},
}
\begin{document}

\maketitle

\begin{abstract}
    Child-centered daylong recordings are essential for studying early language development, but existing speech models trained on clean adult data perform poorly due to acoustic and linguistic differences. We introduce BabyHuBERT, a self-supervised speech model trained on 13,000 hours of multilingual child-centered recordings from 40+ languages. Evaluated on voice type classification, the task of identifying who produces speech and when in child-centered recordings (key child, other children, male, and female adults), BabyHuBERT-VTC achieves F1-scores from 55.0\% to 76.1\% across six corpora, consistently outperforming W2V2-LL4300 and HuBERT (pretrained on English daylongs and clean adult speech, respectively). Notable gains include 14.0 and 18.3 absolute F1 points over HuBERT on Vanuatu and Solomon Islands, demonstrating effectiveness on underrepresented languages. We share code and models to support researchers working with child-centered recordings across diverse linguistic contexts.
\end{abstract}

\section{Introduction}
\label{sec:intro}

\begingroup
\renewcommand\thefootnote{}
\footnotetext{%
  \hspace{-1.8em}\fontsize{7.1}{9}\selectfont
  BabyHuBERT: \url{https://huggingface.co/MarvinLvn/BabyHuBERT}\\ 
  BabyHuBERT-VTC: \url{https://github.com/LAAC-LSCP/VTC} \\
  Release materials: \url{https://osf.io/n75rs}
}
\endgroup

Despite decades of advances in automatic speech processing, speech science and technology have had relatively limited impact on language development research due to a fundamental mismatch: existing speech models, trained predominantly on clean adult speech, fail catastrophically on the complex and diverse acoustic environments where children acquire language~\cite{garcia:hal-02417632}. Child-centered daylong recordings, naturalistic audio captured by devices worn by children throughout their daily lives, present a perfect storm of challenges that systematically break existing models: 80\% non-speech content (silence, noise, or environmental sounds), alongside fragmented speech characterized by short vocalizations, overlapping speakers, variable acoustic conditions, and distant or muffled speech~\cite{cristia2024long,peurey2025fifteen}. The children themselves produce speech with higher fundamental frequencies, greater spectral variability, and non-standard pronunciations that further confound adult-trained systems~\cite{Sungbok_1999,Gerosa_2007}. This technological bottleneck has prevented automated analysis at the scale needed for developmental research, where manual annotation is prohibitively expensive, particularly affecting underrepresented languages~\cite{blasi2021systematic}.

We introduce BabyHuBERT, the first self-supervised representation model pre-trained on massive-scale child-centered daylong recordings spanning 13,164 hours across over 40 languages, from widely-studied languages such as English and French to underrepresented languages including Yeli Dnye, Tsimane, and Quechua. Our approach differs fundamentally from existing speech representation models in both training data and scale (see Section~\ref{sec:related}). Our key contributions are threefold: (1) We built the first large-scale multilingual speech model specifically designed for child-centered recordings; (2) We achieve substantial performance improvements on voice type classification\footnote{Voice type classification, as used here, is equivalent to the term ``speaker segmentation'' as introduced by LENA~\cite{xu2008signal}, the first widely-used system for this task.}, with an average F1-score of 66.9\% that approaches human annotator performance (69.8\%) while consistently outperforming existing models across all tested languages; and (3) We demonstrate that domain-specific pre-training on child-centered speech data leads to substantial improvements over adult-only models. We hope BabyHuBERT will provide the community with representations better suited to child-centered recordings, enabling a broad range of downstream tasks that have until now relied only on models trained on adult speech. 

\section{Related Work}
\label{sec:related}

\subsection{Self-supervised learning on child-centered recordings}

Given the scarcity of self-supervised representation models trained on child-centered recordings, previous work has primarily relied on fine-tuning adult-trained models (e.g.,~\cite{rolland2024introduction,medin2025self} for recent automatic speech recognition examples). While these approaches show improvements over adult-trained models applied directly to child speech, they still rely on representations learned from clean adult speech, limiting their effectiveness on the complex acoustic environments of child-centered recordings.

One previous study addressed the challenges of daylong recordings through self-supervised pre-training: Li et al. (2023)~\cite{li23e_interspeech} introduced W2V2-LL4300, a wav2vec 2.0 model~\cite{wav2vec2_2020} pre-trained on 4,300 hours of child-centered recordings from English-learning children under 5 years old. This work achieved substantial improvements over adult-trained models on child speech tasks, showing that domain-specific pre-training could outperform direct fine-tuning of adult-trained models. Zhang et al. (2025)~\cite{zhang_vcm_2025} fine-tuned W2V2-LL4300 on speech maturity classification, confirming the value of using in-domain data for pre-training. However, W2V2-LL4300 remains limited in two critical ways: an English-only focus that does not generalize to the multilingual nature of language development research~\cite{marvin_rev_eng_la_2022}, and a relatively small scale compared to the vast amounts of unannotated data available~\cite{macwhinney_childes, vandam_homebank_2016}.

\subsection{Voice type classification}

Voice Type Classification (VTC) consists of classifying audio segments into speaker categories, and is fundamental to child language acquisition research, enabling automated analysis of who speaks when in naturalistic recordings~\cite{cristia_2021,lavechin2025performance}. Unlike traditional speaker diarization, which produces anonymous labels (e.g., Speaker 1, Speaker 2), VTC assigns developmentally meaningful categories: knowing whether a speaker is an adult or another child is essential for understanding the child's linguistic environment. In this paper, we consider four categories: key child (\class{KCHI}, the child wearing the recording device), other children (\class{OCH}), male adult (\class{MAL}), and female adult (\class{FEM})~\cite{lavechin20_interspeech}. The multi-label design handles the overlapping speech common in naturalistic recordings, where multiple speaker categories can be active simultaneously.

The proprietary \LENA system, built on Gaussian Mixture Models, was the first VTC system and remains widely used today~\cite{xu2008signal}. PyanNet-VTC provided an open-source alternative based on a SincNet architecture trained on the multilingual BabyTrain dataset~\cite{lavechin20_interspeech,sincnnet_2018}. More recently, Whisper-VTC replaced SincNet with frozen Whisper features and established BabyTrain-2025, a 670-hour multilingual dataset annotated with speaker categories~\cite{kunze25_interspeech}. This work also established human-level performance by having a second annotator label the same data, revealing that current systems significantly underperform human annotators and indicating that the VTC task remains far from solved. We build on these efforts and demonstrate that domain-specific pre-training can substantially close the gap between automated systems and human performance.

\section{Methods}
\label{sec:methods}

% 1. dataset curation and pre-processing
\subsection{Datasets}

% ==========
\begin{table*}[h]
    \fontsize{9pt}{9pt}\selectfont
    \centering
    \caption{Breakdown of our pre-training set used to train BabyHuBERT. Datasets with * are available via a scientific archive (Homebank~\cite{vandam_homebank_2016}, Databrary~\cite{simon2015databrary}, or the Language Archive~\cite{skilton2021ticuna}). Total duration refers to the duration of raw audio. Due to continuous multi-hour recordings, approx. 80\% of raw audio is non-speech (noise or silence). Effective duration is the duration of audio used for the pre-training after pre-processing.}
    \label{tab:pretrain-data}
    \begin{tabular}{lllrr}
        \toprule
        Dataset & Location & Languages &  Total duration (h) & Effective duration (h) \\
        \midrule
        Cougar*~\cite{vandam2018cougar}         & USA               & English             & 8234   & 3535           \\
        Timor-leste2022~\cite{timor_leste_22_baranov} & Timor-Leste       & Multilingual (28 languages)  & 6635  & 1838   \\
        Fausey-trio*~\cite{fausey_trio_2018} & USA               & English             & 1907   & 421            \\
        Solomon~\cite{cassar2025-solomon} & Solomon Islands   & Multilingual (12 languages)  & 5484  & 1855   \\
        Bergelson*~\cite{bergelson2017seedlings, berg_daybday_2019} & USA               & English             & 7065   & 2018           \\
        Lucid*~\cite{lucid_rowland_2025} & UK                & English             & 3557   & 1308           \\
        Png2019~\cite{png2019_cristia} & Papua New Guinea  & Mostly Yeli Dnye    & 855    & 325            \\
        Tsimanem2018~\cite{tsimane_m_c_2018_scaff} & Bolivia           & Tsimane             & 740    & 154            \\
        Png2016~\cite{casillas_png2016} & Papua New Guinea  & Mostly Yeli Dnye    & 483    & 245            \\
        Warlaumont*~\cite{warlaumont_homebank_2024} & USA               & Mostly English      & 499    & 165            \\
        Tsimanec2018~\cite{tsimane_m_c_2018_scaff} & Bolivia           & Tsimane             & 802    & 209            \\
        Tseltal2015~\cite{tseltal_casillas_homebank_2017} & Mexico            & Tseltal             & 502    & 201            \\
        Quechua*~\cite{quechua_cychosz_homebank_2018} & Bolivia           & Quechua,Spanish     & 975    & 374            \\
        Ramirez~\cite{ramirez_2024} & Slovenia          & Slovenian           & 537    & 251            \\
        Israel-Haifa~\cite{israel_haifa_natovich} & Israel            & Hebrew              & 149    & 85             \\
        Winnipeg*~\cite{winnipeg_soderstrom_homebank_2016, winnipeg_soderstrom_journal_2016} & USA               & English             & 351    & 105            \\
        PhonSES~\cite{phonses_cristia_2021} & France            & French              & 114    & 29             \\
        Nepal-havron~\cite{nepal_havron_2023}                      & Nepal             & Mixture             & 103    & 38             \\
        Lyon*~  \cite{lyon_homebank_2016, canault2016} & France            & French              & 37     & 8              \\
        \midrule
        Total            & 11 countries      & $>$40 languages     & 39 029 & 13 164         \\
        \bottomrule
    \end{tabular}
\end{table*}

\textbf{\textit{Pre-training set}}: Our pre-training set comprises 19 diverse corpora spanning multiple continents across 40+ languages (see Table \ref{tab:pretrain-data}). These child-centered recordings capture speech from the child wearing the recording device, other children, and adults present in the environment. Some datasets were obtained from scientific archives like HomeBank~\cite{vandam_homebank_2016}, while others were accessed through direct data-sharing agreements with research groups. To prevent data leakage, we exclude all audio files from our pre-training set that are included in our fine-tuning set (see below), ensuring a clean separation between pre-training and evaluation data. This results in a pre-training set comprising 43\% of non-English content, ensuring broad linguistic and sociocultural diversity.

Daylong recordings are collected continuously over multiple hours, meaning about 80\% of raw audio is non-speech. This differs fundamentally from previous pre-training setups. Pre-training on this raw audio would expose the model predominantly to silence and noise rather than the speech representations we want it to learn. We therefore use PyanNet-VTC~\cite{lavechin20_interspeech} to detect and extract speech segments. We pad segments shorter than 2 seconds with surrounding audio to avoid very short isolated chunks, and merge consecutive segments separated by less than 2 seconds of silence into a single chunk (capped at 30 seconds) to provide the model with sufficient context during pre-training. This preprocessing reduces the overall proportion of non-speech content received by the model to around 8\%.

\textbf{\textit{Fine-tuning set}}: To fine-tune on the voice type classification task, we use BabyTrain-2025~\cite{kunze25_interspeech}, a 670-hour multilingual dataset annotated with speaker 
categories, reflecting the class imbalance inherent to naturalistic recordings: \class{KCHI}: 158h, \class{OCH}: 11h, \class{MAL}: 12h, \class{FEM}: 262h. The dataset is partitioned into training, validation, and test sets with an 80/10/10 split using child-disjoint partitioning to ensure no child appears in multiple splits. We use 20 hours of child-centered audio drawn from the ACLEW corpora collected in English-speaking families~\cite{soderstrom2021developing} as a hold-out set, the same as that used in~\cite{lavechin20_interspeech} to allow for comparison.

% 2. pre-training HuBERT
\subsection{Pre-training strategy}

We use HuBERT to leverage its masked prediction approach, which has shown promise in noisy conditions~\cite{wavlm_2022, mhubert147_2024}, a relevant advantage for child-centered recordings. We consider the base architecture due to computational constraints. We choose HuBERT over WavLM primarily due to the availability of well-documented training code, while starting from WavLM features to further leverage its denoising objective. 

Our pre-training follows a two-iteration approach. In HuBERT, discrete pseudo-labels are generated via K-means clustering of speech features and used as targets for the masked prediction objective. For the first iteration (BabyHuBERT-1), we cluster features extracted from WavLM-base-plus\footnote{\url{https://huggingface.co/microsoft/wavlm-base-plus}}, as done in~\cite{wavlm_2022}, instead of the MFCCs used in the original approach~\cite{hubert_2021}. For the second iteration (BabyHuBERT-2), we cluster features from BabyHuBERT-1's transformer, following standard HuBERT methodology, except we use an earlier layer based on empirical validation on the voice type classification task (see Section~\ref{sec:details}).

% 3. Finetuning the trained model on downstream task
\subsection{Fine-tuning strategy}

To fine-tune BabyHuBERT-1 and BabyHuBERT-2 on the VTC task, we add independent binary classification heads for each speaker type, following~\cite{kunze25_interspeech}. This design allows multiple labels to be activated simultaneously and enables each head to specialize on its respective speaker category.  Following~\cite{hubert_2021}, we fine-tune only the transformer layers, while the convolutional layers remain frozen. Each configuration is trained across 10 different random seeds. We select the best model across all training runs based on validation set performance.

\subsection{Baselines and evaluation}

We benchmark against two categories of baselines. The first includes systems specifically designed for VTC: 
\LENA~\cite{xu2008signal} used off-the-shelf, 
PyanNet-VTC~\cite{lavechin20_interspeech}, and 
Whisper-VTC~\cite{kunze25_interspeech}. The second includes general SSL models fine-tuned on BabyTrain-2025, allowing us to examine, albeit with confounding variables, the role of pre-training data: HuBERT pre-trained on clean adult speech~\cite{hubert_2021} and W2V2-LL4300 pre-trained on English daylong recordings~\cite{li23e_interspeech}. A second human annotator serves as the topline (Human 2), providing an upper bound on expected performance given the difficulty of the task.

All models are evaluated using the F1-score, the harmonic mean between precision and recall, as implemented in \texttt{pyannote.metrics}~\cite{bredin2017pyannote}.

\subsection{Implementation details}
\label{sec:details}

\noindent \textit{\textbf{Pre-training}}: We use the HuBERT base architecture~\cite{hubert_2021} and 
follow its standard training procedure, as implemented in torchaudio~\cite{torchaudio_2023}. For BabyHuBERT-1, we extract features from the 6th layer of WavLM-base-plus. For BabyHuBERT-2, we use features from BabyHuBERT-1's 7th transformer layer. We apply K-means clustering to these extracted features using MiniBatchKMeans with 500 clusters~\cite{scikit_learn_2011}. The clustering is performed on 2,500 hours of randomly sampled features from our training data. For HuBERT training, we use a maximum batch size of 175 seconds (85 effective seconds per GPU after bucketing), training for 400k steps on 32 H100 GPUs, following~\cite{mhubert147_2024}. Both BabyHuBERT iterations are trained for approximately 30 hours each, completing 45 and 44 epochs, respectively. 

\noindent \textit{\textbf{Fine-tuning}}: We fine-tune all models with the same hyperparameters. Our classification heads consist of 4 linear layers (one per label) fed with the encoder's last layer representation and a dropout probability of 0.5. We set the initial learning rate to 1e-5, decreasing it by 10 when the validation loss hits a plateau. All models are trained with a batch size of 128 utterances (10 seconds each) until convergence.

% =====================================================
\section{Results}

% 1. VTC finetuning result table
\begin{table}[h]
    \centering
    \fontsize{9pt}{9pt}\selectfont
    \caption{F1-scores (\%) obtained on the hold-out set by HuBERT-base and HuBERT-large~\cite{hubert_2021}, W2V2-LL4300 \cite{li23e_interspeech}, and BabyHuBERT iteration 1 and 2. All models are fine-tuned on BabyTrain-2025. Due to space constraints, standard deviation is reported only for the F1-score averaged across speaker categories (Ave.). Best performance is shown in bold.}
    \label{tab:vtc-ff-vs-frozen}
    \setlength{\tabcolsep}{4.3pt}
    \begin{tabular}{@{}lccccc@{}}
        \toprule
        & \multicolumn{5}{c}{F1-scores (\%) $\uparrow$} \\
        Model & \class{KCHI} & \class{OCH} & \class{MAL} & \class{FEM} & Ave. \\
        \midrule
        HuBERT base  & 62.7 & 29.1 & 47.0 & 64.1 & $50.7 \pm 1.7$ \\
        HuBERT large  & 63.5 & 27.7 & 42.7 & 62.6 & $49.1 \pm 3.6$ \\
        W2V2-LL4300 & 68.1 & 41.0 & 55.4 & 69.1 & $58.4 \pm 1.1$ \\
        BabyHuBERT 1  & 69.7 & 47.0 & 63.0 & 74.1 & $63.5 \pm 1.6$ \\
        % BabyHuBERT 2 & \textbf{70.0} & \textbf{47.9} & \textbf{63.5} & \textbf{74.7} & $\mathbf{64.0 \pm 1.0}$ \\
        BabyHuBERT 2 & \textbf{70.7} & \textbf{46.9} & \textbf{64.3} & \textbf{74.9} & $\mathbf{64.2 \pm 1.0}$ \\
        \bottomrule
    \end{tabular}
\end{table}

\noindent\textit{\textbf{Fine-tuning experiments}}: Table~\ref{tab:vtc-ff-vs-frozen} reports the hold-out set performance for all fine-tuned models. HuBERT base and large, pre-trained on clean adult English speech, achieve the lowest performance (average F1-score of $50.7 \pm 1.7\%$ and $49.1 \pm 3.6 \%$, respectively). The substantial performance gain observed with W2V2-LL4300 (+7.7\% average F1 over HuBERT base) likely stems from its pre-training on child-centered recordings collected in English-speaking families. Unlike controlled adult speech corpora, such recordings expose the model to naturalistic child and child-directed speech as well as the challenging acoustic conditions typical of daylong recordings.

Both BabyHuBERT iterations substantially outperform W2V2-LL4300 ($63.5 \pm 1.6 \%$ and $64.2 \pm 1.0\%$ vs.$58.4 \pm 1.1\%$), likely because BabyHuBERT benefits from both a larger pre-training corpus and greater linguistic diversity. This exposes the model to a broader range of child vocal patterns and acoustic conditions beyond those found in English-speaking households. Both iterations perform similarly, suggesting that the additional clustering iteration provides limited additional gains. Given that BabyHuBERT-2 achieves the highest F1-scores across all speaker categories, we select it for all subsequent experiments (hereinafter referred to as BabyHuBERT-VTC).

\begin{table}[h]
    \centering
    \fontsize{9pt}{9pt}\selectfont
    \setlength{\tabcolsep}{4.3pt}
    \caption{F1-scores (\%) obtained on the hold-out set by \LENA, PyanNet-VTC \cite{lavechin20_interspeech}, Whisper-VTC \cite{kunze25_interspeech}, the best BabyHuBERT-VTC model and compare it to a second human annotator (Human 2). Best performance is shown in bold.}
    \label{tab:vtc-sota}
    \begin{tabular}{@{}lccccc@{}}
        \toprule
        & \multicolumn{5}{c}{F1-scores (\%) $\uparrow$} \\
        Model & \class{KCHI} & \class{OCH} & \class{MAL} & \class{FEM} & Ave. \\
        \midrule
        \LENA \cite{xu2008signal}   & 54.9  & 28.5 & 37.2 & 42.6 & 40.8 \\
        PyanNet-VTC \cite{lavechin20_interspeech} & 68.2 & 30.5 & 41.2 & 63.7 & 50.9 \\
        Whisper-VTC \cite{kunze25_interspeech} & \textbf{68.4} & 20.6 & 56.7 & 68.9 & 53.6 \\
        % BabyHuBERT-VTC (ours)  & \textbf{70.0} & \textbf{50.9} & \textbf{65.1} & \textbf{74.3} & \textbf{65.1} \\
        BabyHuBERT-VTC (ours)  & 67.1 & \textbf{56.1} & \textbf{68.8} & \textbf{75.5} & \textbf{66.9} \\
        Second human annotator & 79.7 & 60.4 & 67.6 & 71.5 & 69.8 \\
        \bottomrule
    \end{tabular}
\end{table}

\noindent\textit{\textbf{Comparison with existing models}}: In Table~\ref{tab:vtc-sota}, we compare the performance of our best model, BabyHuBERT-VTC, against existing baselines. We also report the performance of a second human annotator as a topline, reflecting the inherent ambiguity of the task and providing an upper bound that accounts for disagreement between annotators. Results demonstrate that BabyHuBERT-VTC substantially outperforms existing systems on the hold-out set. Our best model achieves 66.9\% average F1-score, representing 13.3 and 16.0 absolute percentage point improvements over Whisper-VTC and PyanNet-VTC, respectively. The model shows notable progress on the challenging \class{OCH} class, achieving 56.1\% F1 compared to much lower performance from existing systems. \class{OCH} is particularly difficult as other children's voices share many acoustic properties with the key child's voice, making the two classes hard to discriminate.  Comparison with a second human annotator reveals that BabyHuBERT approaches human-level performance, achieving 66.9\% average F1-score compared to the second annotator's 69.8\%, representing a gap of only 2.9 percentage points. 

\begin{figure}[h]
    \centering
    \includegraphics[width=\linewidth]{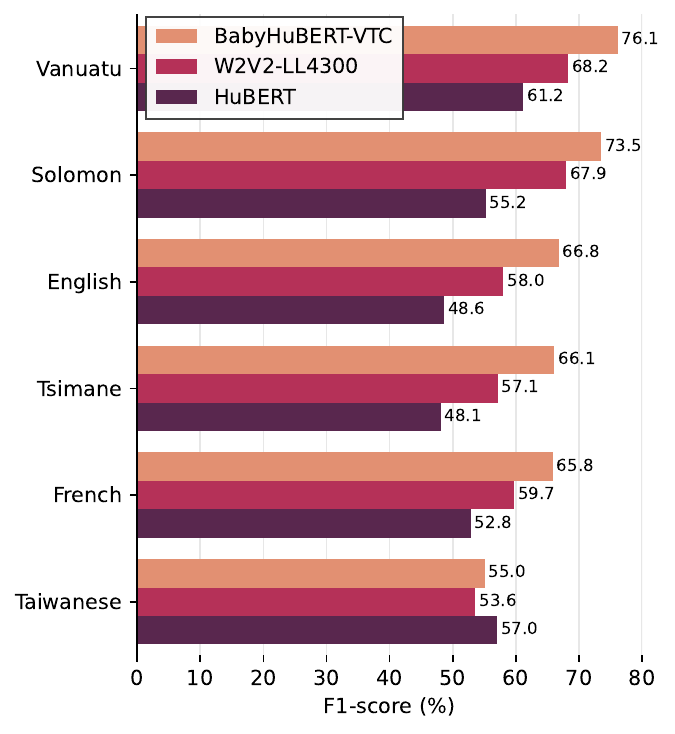}
    \caption{Comparison of the performance obtained by BabyHuBERT-VTC, W2V2-LL4300 and HuBERT (all fine-tuned on BabyTrain-2025) across corpora of the test set.}
    \label{fig:per-per-lang-per-model}
\end{figure}

\noindent{\textbf{\textit{Performance across datasets}}}: While using a hold-out set facilitates comparison with previous papers, we turn to BabyTrain-2025's test set to assess the models' ability to handle diverse languages. Figure~\ref{fig:per-per-lang-per-model} shows that W2V2-LL4300 consistently outperforms standard HuBERT (showing the importance of in-domain pre-training) but falls short of BabyHuBERT across all conditions, reinforcing the value of large-scale multilingual training over English-only child data. Performance varies significantly across datasets, with English achieving a mid-range score despite being the most represented language in our training data (57\%) and the only one in W2V2-LL4300's pre-training set. This suggests that linguistic background may not be a major determinant of performance. Moreover, all models achieve their highest performance in Vanuatu and the Solomon Islands, which are highly multilingual. We also observe that corpora featuring larger families in small-scale societies (Vanuatu, the Solomon Islands, Tsimane) are spread out, suggesting number and identity of speakers may not be a determining factor either. Finally, we note that variation in performance across corpora seems similar across models, suggesting that they are all picking up on similar difficulties despite wide differences in pre-training data.

% =====================================================
\section{Discussion}
% findings
Child-centered recordings represent a formidable challenge for speech science and technology, and the scarcity of human annotations has held back progress. The difficulty of such data is obvious when considering our human annotator topline: when comparing two humans' decisions on who speaks when (key child, other children, female adults, male adults), we arrive at an F1-score of 69.8\%. Building on previous work~\cite{kunze25_interspeech,li23e_interspeech}, we demonstrate the value of domain-specific pre-training on child-centered recordings: our best model achieves an 13.3\% absolute F1-score improvement over the previous state of the art, substantially closing the gap to human performance.

Regarding future work, it is clear that pre-training is crucial, particularly to achieve reasonable performance in under-represented classes (other children and male adults). Comparison against W2V2-LL4300 also suggests that increasing size and/or diversity in the pre-training dataset yields the largest gains for these same two classes. In particular, we anticipate that the 56.1\% F1-score on other children's speech (a 25.6 absolute point improvement over the previous state of the art) will enable new research on peer interaction, which is gaining attention given the previously unrecognized importance of sibling speech for development~\cite{loukatou2022child}. 

Due to the computational cost of self-supervised pre-training, we were limited in our ability to explore different hyperparameter configurations (number of clusters, number of training steps, etc.) during BabyHuBERT's pre-training, and systematic optimization of these parameters could yield further improvements. That said, the 60.4\% F1-score achieved by the second human annotator on other children's speech suggests that even manual annotation of this class is inherently difficult, likely due to the limitations of single-microphone recordings. We hope future work explores alternative hardware setups, such as multiple recording devices tracking speaker positions~\cite{perry2024putting} or contact microphones, which could provide richer information and enable more reliable ground-truth annotation, particularly to discriminate the key child from other children.

While BabyHuBERT represents a significant technical advance, we acknowledge important ethical considerations around sharing models trained on sensitive data involving children from under-represented communities. We therefore release BabyHuBERT under a custom license informed by an independent ethics assessment covering risks such as participants' consent, indigenous data sovereignty, privacy, and possible misuse. The license prohibits commercial use and surveillance, requires reporting of misuse, and ensures that any derivative model inherits the same conditions. BabyHuBERT-VTC and associated code are released under the same license. 

Beyond voice type classification, we expect BabyHuBERT to benefit a broad range of downstream tasks in child language research, including addressee classification or phoneme recognition, particularly for underrepresented languages where adult-trained models fall short.

\newpage
\section{Acknowledgments}
%Hidden to protect anonymity.
This work was performed using HPC resources from GENCI-IDRIS (2024-AD01101545 and 2025-AD011016414) and was supported in part by the Agence Nationale pour la Recherche (ANR-17-EURE-0017 Frontcog, ANR10-IDEX-0001-02 PSL*). ED, TC, AC, and TK acknowledge funding from the European Research Council (InfantSimulator-101142705 and ExELang-101001095). ML acknowledges funding from Simons Foundation International (034070-00033) and MP from Agence de l’Innovation de Défense. Views and opinions expressed are those of the authors only and do not necessarily reflect those of grant sources.

%commented for arxiv
%\section{Generative AI Use Disclosure}
%Generative AI tools were used for minor language editing and proofreading of this manuscript. All technical content, experimental design, analysis, and conclusions are entirely the authors' own.

%\clearpage
\begingroup
    \hypersetup{hidelinks}
    \bibliographystyle{IEEEtran}
    \bibliography{refs}
\endgroup
\end{document}